\documentclass[prl,aps,showpacs,twocolumn,floatfix]{revtex4}
\usepackage{epsfig} \usepackage{graphics} \usepackage{bm}
\usepackage{amssymb}
\usepackage{graphicx}
\begin{document}

\preprint{Lebed-PRB}

\title{Possible Existence of Superconductivity in the Quasi-One-Dimensional
Conductor Li$_{0.9}$Mo${_6}$O$_{17}$ at Ultrahigh Magnetic Fields,
$H \geq 45 \ T$. }

\author{A.G. Lebed$^*$}

\affiliation{Department of Physics, University of Arizona, 1118 E.
4-th Street, Tucson, AZ 85721, USA}

\begin{abstract}
We derive a so-called gap equation for superconductivity in a
quasi-one-dimensional (Q1D) layered conductor in a magnetic field,
which is parallel to both its conducting plane and its conducting
axis. This equation demonstrates that the orbital destructive
effects against superconductivity cannot destroy it, if the
perpendicular to the plane coherence length is less than the
inter-plane distance. On the basis of our results, we suggest
arguments that some triplet superconducting phase was indeed
discovered in ultrahigh magnetic fields in the Q1D superconductor
Li$_{0.9}$Mo$_6$O$_{17}$ [see Xiaofeng Xu et al., Phys. Rev.
Lett., $\bf{102}$, 206602 (2009).] We also discuss a possibility
that the above-mentioned superconducting phase can be stable at
arbitrary high magnetic fields.
\end{abstract}

\pacs{74.20.Rp, 74.25.Op, 74.70.Kn}

\maketitle

Historically, in quasi-one-dimensional (Q1D) compounds,
superconductivity was first discovered in the organic conductors
with chemical family (TMTSF)$_2$X, where X=PF$_6$, ClO$_4$, etc.
[1,2]. Note that, from the beginning, (TMTSF)$_2$X superconductors
were considered as good candidates for triplet superconducting
pairing [3-7]. Nevertheless, more recently it has been
found that the superconductor (TMTSF)$_2$ClO$_4$ is actually a
singlet $d$-wave like one [8-10], whereas the superconductor
(TMTSF)$_2$PF$_6$ can still be considerate as a candidate for
triplet superconductivity [5-7,11] or for singlet-triplet mixed
superconducting phase [12]. The reason for the latter is that the
upper critical magnetic fields in the (TMTSF)$_2$PF$_6$ strongly
exceed [5-7] the Clogston-Chandrasekhar paramagnetic limit [13]
and that the Knight shift in superconducting phase has not been
found. Another Q1D superconductor, Li$_{0.9}$Mo${_6}$O$_{17}$, has
recently attracted much attention since there has been shown [14]
that its upper critical field, parallel to the conducting axis,
five times exceeds the Clogston-Chandrasekhar paramagnetic limit.
On this basis, there has been recently suggested [14-16] that the
superconductor Li$_{0.9}$Mo${_6}$O$_{17}$ is a triplet one with
spin part of the order parameter being unsensitive to the Pauli
destructive effects against superconductivity. Very recently
discovered superconductors K$_2$Cr$_3$As$_3$ and
Rb$_2$Cr$_3$As$_3$ [17-19] have added  potential candidates for a
triplet superconducting pairing in Q1D conductors [17,18], although
in Ref.[20] the exceeding of the paramagnetic limit in the
K$_2$Cr$_3$As$_3$ is prescribed to the possible many bands effects.

Note that superconductivity in the Li$_{0.9}$Mo${_6}$O$_{17}$ in
Ref.[14] is observed under experimental conditions, where there is
an increase of resistivity near the superconducting transition
temperature. In Ref. [21], this effect was prescribed to the
coexistence of superconductivity with some paramagnetically
limited phase - presumably charge-density-wave (CDW). Moreover, it
was shown that at very high magnetic fields, $H \simeq 45 \ T$,
where the possible CDW is destroyed by the Pauli spin-splitting
effects, some extremely low conductive state appears at $T_c
\simeq 10 \ K$ [21]. In Ref.[21], it was put forward a hypothesis
that it may be a triplet superconducting phase. The goal of our
paper is to show that parallel to the conducting plane and
conducting axis magnetic field can preserve triplet
superconductivity if the coherence length, perpendicular to the
plane, is less than the inter-plane distance. In this context, we
pay attention that the opposite case is considered by us in
Ref.[15], where the parallel magnetic field destroys
superconductivity. We discuss how our current theoretical results
support the hypothesis [21] about the existence of ultrahigh
magnetic field triplet superconductivity as well as give a hint
that this superconducting phase may survive even at higher
magnetic fields in the Li$_{0.9}$Mo${_6}$O$_{17}$. As a result, we
suggest to perform more experiments in steady magnetic fields of
the order of $H \simeq 50 \ T$ and in pulsed magnetic fields of $H
> 50 \ T$.

Let us consider a tight binding model for a Q1D layered electron
spectrum:
\begin{equation}
\epsilon({\bf
p})=-2t_x\cos(p_xa_x)-2t_y\cos(p_ya_y)-2t_z\cos(p_za_z),
\end{equation}
where $\hbar \equiv 1$ and $t_x \gg t_y \gg t_z$. Note that the
latter inequality allows us to linearize electron spectrum near
two sheets of Q1D Fermi surface (FS), which can be expressed as:
\begin{equation}
p_x(p_y)=\pm p_F \pm \frac{2t_y}{v_F}\cos(p_ya_y),
\end{equation}
where $\pm$ stands for the right (left) sheet of the FS. Let us
introduce in-plane velocity component, which is perpendicular to
the conducting axis,
\begin{equation}
v_y(p_x) = \partial \epsilon({\bf p})/\partial p_y =
2t_ya_y\sin[p_y(p_x)a_y].
\end{equation}
As mentioned above the energy spectrum (1) can be linearized near
the left and right sheets of the FS. Measured with respect to the
Fermi energy, ${\delta \epsilon= \epsilon -\epsilon_F}$, the
linearized electron spectrum (1) takes the form
\begin{equation}
\delta \epsilon^{\pm}(\textbf{p})=\pm v_y(p_y)
\left[p_y-p_y^{\pm}(p_x)\right] -2t_z\cos(p_za_z).
\end{equation}
In a magnetic field, parallel to the conducting ${\bf x}$ axis, it
is convenient to choose the corresponding vector potential in the
following form:
\begin{equation}
{\bf A}=(0,0,Hy), \ \ \ {\bf H}=(H,0,0).
\end{equation}

To obtain Hamiltonian in the magnetic field (5) for the
electron spectrum (4), the so-called Peierls substitution method
is used:

\begin{equation}
p_y-p_y^{\pm}(p_x)\rightarrow -i\frac{\partial }{\partial y} \  \
\ \mathrm{and}  \  \  \  p_z \rightarrow p_z - \frac{e}{c}A_z ,
\end{equation}
where $A_z = Hy$ is the $z$ component of the vector potential; $e$ is the electron charge and $c$ is the
velocity of light, correspondingly. As
a result, the following Hamiltonian is obtained:

\begin{eqnarray}
&\delta\hat{\epsilon}^{(\pm)}=\mp
iv_y[p_y^{\pm}(p_x)]\frac{\partial}{\partial y} -
2t_z\cos\left(p_za_z- \frac{\omega_z}{v_F}y\right), \nonumber\\
&\omega_z=eHa_zv_F/c ,
\end{eqnarray}
where we do not take into account an electron spin, since below we
consider only those triplet superconducting phases, which are not
sensitive to the Pauli spin-splitting effects in a magnetic field.
The total orbital electron wave functions, which are both
eigenfunctions of energy and parallel momentum with the
eigenvalues being $\delta \epsilon$ and $p_x$, can be written in
the factorized form

\begin{equation}
\Psi^{\pm}_{\delta \epsilon, p_x}(x,y,p_z)=e^{\pm ip_x x}\,e^{\pm
ip_y^{\pm}(p_x)y} \ \psi^{\pm}_{\delta \epsilon, p_x}(y,p_z).
\end{equation}
Here, wave functions in the mixed representation,
$\psi^{\pm}_{\delta \epsilon, p_x}(y,p_z)$, are the solutions of
the following Schr\"{o}dinger-like equation: $\delta
\hat{\epsilon}^{\pm} \, \psi^{\pm}_{\delta\epsilon, p_x}(y,p_z) =
\delta \epsilon \, \psi^{\pm}_{\delta \epsilon, p_x}(y,p_z)$,
which can be written as

\begin{eqnarray}
&\mp i v_y[p_y^{\pm}(p_x)]\frac{\partial \psi^{\pm}_{\delta
\epsilon, p_x}(y,p_z)}{\partial y} =\biggl[ \delta \epsilon
+2t_z \cos\biggl( p_za_z -\frac{\omega_z}{v_F}y \biggl)\biggl]
\nonumber\\
&\times \psi^{\pm}_{\delta \epsilon, p_x}(y,p_z).
\end{eqnarray}
It is important that Eq.(9) can be analytically solved:

\begin{eqnarray}
&\psi^{\pm}_{\delta \epsilon, p_x}(y,p_z)= \exp\biggl[\pm i
\frac{2t_z}{v_y[p_y^{\pm}(p_x)]}\int_0^y
\cos\biggl(p_za_z-\frac{\omega_z}{v_F}y'\biggl)\mathrm{d}y'\biggl]
\nonumber\\
&\times \exp\biggl(\pm i \frac{\delta \epsilon \,
y}{v_y[p_y^{\pm}(p_x)]}\biggl).
\end{eqnarray}
Therefore, the total normalized wave functions (8) are:

\begin{eqnarray}
&{\Psi^{\pm}_{\delta \epsilon, p_x}(x,y,p_z)= \frac{e^{\pm ip_x
x}e^{\pm i p_y^{\pm}(p_x) y}}{\sqrt{2\pi |v_y[p_y^{\pm}(p_x)]|}}
\exp\left(\pm i
\frac{\delta \epsilon \ y}{v_y[p_y^{\pm}(p_x)]}\right) } \nonumber\\
&\times \exp\left[\pm i \frac{2t_z}{v_y[p_y^{\pm}(p_x)]}
 \int_0^y\cos{\left(p_za_z-\frac{\omega_z}{v_F}y'\right)}\mathrm{d}y'\right].
\end{eqnarray}
Since the wave functions are known (11), we can calculate the
finite temperature Green's functions by means of the standard
procedure [22]:

\begin{equation}
G^{\pm}_{i\omega_n}(\textbf{r},\textbf{r}')= \sum_{\delta
\epsilon}\frac{[\Psi^{\pm}_{\delta \epsilon}(\textbf{r}')]^*
\Psi^{\pm}_{\delta \epsilon}(\textbf{r})}{i \omega_n- \delta
\epsilon},
\end{equation}
where $\omega_n= 2 \pi T (n+1/2)$ is the so-called Matsubara's
frequency.

In this paper, we consider the simplest triplet scenario of
superconductivity in the Li$_{0.9}$Mo$_6$O$_{17}$, which
corresponds to equal spin pairing with fully gapped Fermi surface
[15]:
\begin{equation}
\hat \Delta(p_x,y) = \hat I  \ sgn(p_x) \ \Delta(y) ,
\end{equation}
where $\hat I$ is a unit matrix in spin-space, $sgn(p_x)$ takes
into account the fact that the triplet superconducting order
parameter (13) changes its sign on two slightly corrugated sheets
of the Q1D FS, $\Delta(y)$ is due the orbital effects
against superconductivity in a magnetic field. It is important
that the triplet order parameter (13) is not sensitive to the Pauli
spin-splitting destructive effects in a magnetic field.

Let us make use of the Gor'kov's equations for unconventional
superconductivity [4,23-25] to derive the so-called gap equation for
superconducting order parameter $\Delta(y)$ in Eq.(13). As a
result of lengthly but straightforward calculations, we
obtain:
\begin{eqnarray}
\Delta(y) = g  \biggl< \int_{|y-y_1| > \frac{|v_y(p_y)|}{\Omega}}
\frac{2 \pi T dy_1}{v_y(p_y) \sinh \bigl[ \frac{2 \pi T
|y-y_1|}{v_y(p_y)} \bigl]} \ \Delta (y_1)
\nonumber\\
\times J_0 \biggl\{ \frac{8 t_z v_F}{\omega_z v_y(p_y)} \sin \biggl[
\frac{\omega_z (y-y_1)}{2v_F} \bigg] \sin \biggl[ \frac{\omega_z
(y+y_1)}{2v_F} \bigg] \biggl\} \biggl>_{p_y}  ,
\end{eqnarray}
where $<...>_{p_y}$ stands for the averaging procedure over
momentum component $p_y$, $g$ is the effective electron coupling
constant, and $\Omega$ is the cutoff energy. We pay attention that
Eq.(14) is very general and was derived by us earlier in Refs.
[10,15]. Indeed, Eq.(14) contains several different limiting cases
for the different regions of its parameters. First of all, for low
magnetic fields, it reduces to the Ginzburg-Landau equation [26]
and this limiting case will be considered below as a part of more
general Lawrence-Doniach-like description [27,28]. At extremely
high magnetic fields, where
\begin{equation}
\omega_z(H) \geq \frac{8 t_z v_F}{v^0_y} \ , \ \ v^0_y = 2 t_y a_y ,
\end{equation}
or very low temperatures,
\begin{equation}
T \leq \tilde T(H) \simeq \frac{\omega_z(H)v^0_y}{2 \pi^2 v_F}  ,
\end{equation}
quantum effects of electron motion in a magnetic field become
strong [2] and novel the so-called Reentrant Superconductivity
phase may appear [4,25,29]. If the inequalities (15),(16) are not
fulfilled and, therefore,
\begin{equation}
\omega_z(H) \ll \frac{8 t_z v_F}{v^0_y} \ , \ \ v^0_y = 2 t_y a_y ,
\end{equation}
\begin{equation}
T \gg \tilde T(H) \simeq \frac{\omega_z(H)v^0_y}{2 \pi^2 v_F}  ,
\end{equation}
then two different physical pictures are possible: (a) anisotropic
3D superconductivity, where superconductivity disappear at high enough
magnetic fields, (b) the Lawrence-Doniach-like description, where, as shown
below, triplet superconductivity (13) survives at arbitrary high magnetic field.
The case (a) corresponds to situation, where perpendicular to the plane
coherence length,
 $\xi_{z}$, is bigger than the inter-plane distance, $a_z$, and has been
 already considered in Ref.[15].
The case (b), corresponding to situation, where the perpendicular
coherence length is less than the inter-plane distance,
\begin{equation}
\xi^2_{z} \ll a^2_z ,
\end{equation}
where according to Ref. [15],
\begin{equation}
\xi_z = \frac{\sqrt{7 \zeta(3)} t_z a_z}{2 \sqrt{2} \pi T_c},
\end{equation}
is considered below. Here, $\zeta(x)$ is the so-called Riemann
zeta-function.

It is possible to show that under conditions (17),(18) the
integral Eq.(14) can be rewritten in the following way [15]:
\begin{eqnarray}
\Delta(y) = g  \biggl< \int_{|y-y_1| > \frac{|v_y(p_y)|}{\Omega}}
\frac{2 \pi T dy_1}{v_y(p_y) \sinh \bigl[ \frac{2 \pi T
|y-y_1|}{v_y(p_y)} \bigl]}
\nonumber\\
\times J_0 \biggl\{ \frac{4 t_z (y-y_1)}{v_y(p_y)}
\sin \biggl[ \frac{\omega_z
(y+y_1)}{2v_F} \bigg] \biggl\} \Delta(y_1) \biggl>_{p_y}  .
\end{eqnarray}
Then, if inequality (19) is also fulfilled, it is easy to make
sure that the integral (21) converges at small values of variable
$y -y_1 \sim v^0_y/(2\pi T_c)$, where $T_c$ is superconducting
transition temperature in the absence of a magnetic field.
Therefore, we can further simplify the Bessel function,
\begin{eqnarray}
&J_0 \biggl\{ \frac{4 t_z (y-y_1)}{v_y(p_y)}
\sin \biggl[ \frac{\omega_z
(y+y_1)}{2v_F} \bigg] \biggl\}  \nonumber\\
&\simeq 1 - \frac{2 t^2_z (y-y_1)^2}{v^2_y(p_y)} +
\frac{2t^2_z(y-y_1)^2}{v^2_y(p_y)} \ \cos \biggl(\frac{2 \omega_z
y}{v_F} \biggl),
\end{eqnarray}
and expand superconducting order parameter $\Delta(y_1)$ into a series
of small parameter $y-y_1$:
\begin{equation}
\Delta(y_1) \simeq \Delta(y) + (y_1-y) \frac{d \Delta(y)}{d y} + \frac{1}{2}
(y_1-y)^2 \frac{d^2 \Delta}{dy^2}.
\end{equation}

As a result, after some lengthly but straightforward calculations,
we obtain:
\begin{equation}
-\xi^2_y \frac{d^2 \Delta}{dy^2} +2
\biggl(\frac{\xi_z}{a_z}\biggl)^2 \biggl[1-\cos\biggl(\frac{2
\omega_z y}{v_F}\biggl) \biggl] + \biggl( \frac{T-T_c}{T_c} \biggl)
\Delta(y)=0 ,
\end{equation}
where
\begin{equation}
\xi_y = \frac{\sqrt{ 7 \zeta(3)} t_y a_y}{2\sqrt{2}\pi T_c}.
\end{equation}
Note that to derive Eq.(24) from Eqs.(21)-(23), we have used the
following relationships:
\begin{equation}
1 = g \int^{\infty}_{v_F/\Omega}\frac{2 \pi T_c dx}{v_F \sinh
\biggl(\frac{2 \pi T_c x}{v_F}\biggl)}
\end{equation}
and [30]
\begin{equation}
\int^{\infty}_0 \frac{x^2 dx}{\sinh (x)} = \frac{7 \zeta(3)}{2},
\end{equation}
It is important that, from mathematical point of view, the
obtained equation (24) is similar to equations, obtained in the
Lawrence-Doniach model [27,28]. Nevertheless, we pay attention
that there are two main important differences between our results
and the results of Refs.[27,28]. First of all, we consider Q1D
superconductor in a magnetic field, parallel to its conducting
axis, whereas in Refs.[27,28] only a Q2D superconductor is
studied. Second, we have derived Eq.(24) from very general Eq.(14)
- not from phenomenological Lawrence-Doniach model [27,28].

Note that Mathieu equation (24) is studied in details in the
existing literature (see, for example, Refs.[27,28].) In the near
vicinity of superconducting transition temperature, $(T_c-T)/T_c
\ll (\xi_z/a_z)^2$, Eqs.(24) leads to the Ginzburg-Landau formula
for the upper critical magnetic field,
\begin{equation}
H^x_{c2}(T) = \frac{\phi_0}{2 \pi \xi_y \xi_z} \biggl( \frac{T_c-T}{T_c} \biggl)
= \frac{4  \pi^2 c T^2_c}{7 \zeta(3) e t_y t_z a_y
a_z} \biggl( \frac{T_c-T}{T_c} \biggl),
\end{equation}
where $\phi_0 = \pi c/e$ is the magnetic flux quantum. Note that the
Ginzburg-Landau case corresponds to relatively low magnetic
fields,
 \begin{equation}
 H \ll H^* = \frac{\phi_0 \xi_z}{2 \pi a^2_z \xi_y}
 \end{equation}
On the other hand, if magnetic field is high enough,
\begin{equation}
 H \geq  H^*,
 \end{equation}
then a magnetic field fully penetrates between the conducting layers.
Mathematically, this corresponds to the case, where $\cos(...)$ potential
averages in the Schr\"{o}dinger-like equation (24) and, therefore,
in this case
\begin{equation}
\lim_{H/H^* \rightarrow \infty} T_c(H) \rightarrow T^*_c = T_c[1 -
2 (\xi_z/a_z)^2]
>0 .
\end{equation}
The overall dependence of superconducting transition temperature
on a magnetic field is sketched in Fig.1, where it is clear that,
under condition (19), the triplet superconducting phase (13) is
stable at arbitrary high magnetic fields.

\begin{figure}[t]
\centering
\includegraphics[width=0.5\textwidth]{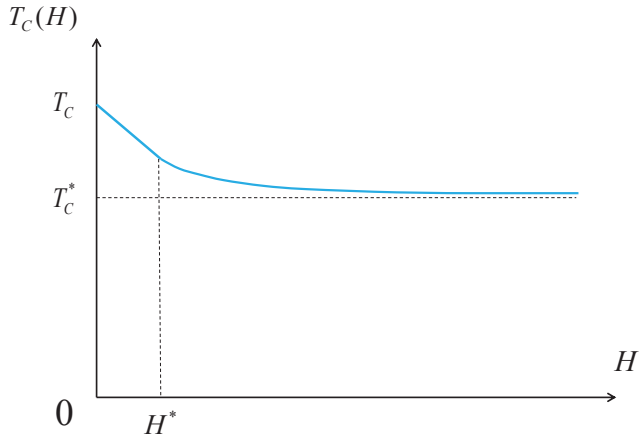}
\caption{Magnetic field dependence of superconducting transition temperature,
$T_c(H)$, derived from Eq.(24), is sketched by solid line. Characteristic
magnetic field, $H^*$, and temperature, $T^*_c$, are given by Eq.(29) and
Eq.(31), respectively.}
\end{figure}

Let us demonstrate that inequalities (17)-(19), used in the paper
for the derivation of Eq.(24), are likely fulfilled in the Q1D
conductor $Li_{0.9}Mo_6O_{17}$ at experimental parameters of
Ref.[21], where $H \simeq 45 \ T$ and $T_c \simeq 10 \ K$. Indeed,
if we make use of the following electron band parameters,
evaluated by us before in Ref.[15] (see Table I), $t_y \simeq 41 \
K$ and $t_z \simeq 14 \ K$, and known values of the crystalline
lattice parameters, $a_z = 9.5 \ \AA$ and $a_y = 12.7 \ \AA$, then
we obtain that Eq.(17) is valid at $H \ll 700 \ T$. Moreover,
using the value of $v^0_y = 1.4 \times 10^6 \ cm/s$ [15], we find
that $\tilde T \simeq 0.3 \ K$. So Eqs.(17),(18) are justified
very well. Let us estimate the parameter $(\xi_z/a_z)^2$, which
has to be small for validity of Eq.(24). To this end, we use
Eq.(20) and the value $T_c \simeq 10 \ K$. As a result, we obtain
the following values: $(\xi_z/a_z)^2 \simeq 0.2$ and
\begin{equation}
T_c^* \simeq 6 \ K ,
\end{equation}
which is in general agreement with the experiment [21]. It is
interesting also to define how far from the Ginzburg-Landau region
the possible superconductivity exists at $H \simeq 45 \ T$ in the
Li$_{0.9}$Mo$_6$O$_{17}$ [21]. Our estimation by means of Eq.(29)
shows that the parameter $H^*$ is equal to $H^* \simeq 95 \ T$,
therefore, experimental superconductivity probably corresponds to
the Ginzburg-Landau region or exists close to it (see Fig.1).

To summarize, we have derived very general Eq.(14) to describe
behavior of superconductivity in a layered Q1D conductor in a
parallel magnetic field. We have considered in detail the
particular case, where Eq.(14) can be reduced to the Mathieu
equation (24), which solution is known. We have considered the
case, where perpendicular to the conducting plane coherence
length, $\xi_z$, is much less than the inter-plane distance,
$a_z$, in the Mathieu equation (24), and showed that, in this
case, superconductivity is stable at arbitrary high magnetic
fields (see Fig.1). We have calculated the corresponding
parameters for the Q1D superconductor $Li_{0.9}Mo_6O_{17}$, using
the results of Refs.[15,21], and demonstrated that our theory is
very likely applicable for it at the experimental field $H \simeq
45 \ T$ [21]. Therefore, we conclude that very low resistive
state, discovered in Ref.[21] and tentatively prescribed to
superconductivity, is very likely indeed superconducting phase. We
stress that very low conducting phase in Ref.[21] is observed at
$T_c \leq 10 \ T$ and magnetic field $H \simeq 45 \ T$, which is
more than two times high than the the Clogston-Chandrasekhar
paramagnetic limit for singlet superconductivity [13], therefore,
the possible superconductivity, discovered in Ref.[21], is
probably of a triplet nature. Nevertheless, at this point, we
cannot completely exclude a singlet nature of the possible
superconducting phase, since, in Q1D superconductors, singlet
superconductivity can exist above the Clogston-Chandrasekhar
paramagnetic limit in the form of Larkin-Ovchinnikov-Fulde-Ferrell
phase [31-33,10,11]. More experimental works are needed to justify
the nature of possible superconductivity at $H \simeq 45 \ T$ as
well as to explore the possibility, suggested in this paper, that
superconductivity can be stable at arbitrary high magnetic fields.

We are thankful to N.N. Bagmet (Lebed), N.E. Hussey, and J.
Singleton for useful discussions. This work was partially
supported by the NSF under Grant No DMR-1104512.

$^*$Also at: L.D. Landau Institute for Theoretical Physics, RAS, 2
Kosygina Street, Moscow 117334, Russia.

\end{document}